\documentclass[12pt,dvips]{article}
\usepackage{amscd,verbatim}
\usepackage[all]{xy}
\usepackage{graphicx}

\usepackage{amsmath}
\usepackage[psamsfonts]{amssymb}
\usepackage{amsthm}
\usepackage{euscript}

\usepackage{latexsym}

\renewcommand{\(}{\begin{equation}}
\renewcommand{\)}{end{equation} \vspace{-.05in}\linebreak}

\newcounter{saveeqn}
\newcounter{savealpheqn}

\newcommand{\alpheqn}{\setcounter{saveeqn}{\value{equation}}%
  \stepcounter{saveeqn}\setcounter{equation}{0}%
  \renewcommand{\theequation}{\mbox{\arabic{section}.\arabic{saveeqn}
\alph{equation}}}
  \renewcommand{\)}{\end{equation}}}
\def\part#1{\frac{\partial}{\partial{#1}}}%
\def\group#1{\refstepcounter{equation}\setcounter{saveeqn}{\value{equati
on}}%
  \label{#1}\setcounter{equation}{0}%
\renewcommand{\theequation}{\mbox{\arabic{section}.\arabic{saveeqn}
\alph{equation}}}
  \renewcommand{\)}{\end{equation}}}
\newcommand{\reseteqn}{\setcounter{equation}{\value{saveeqn}}%
  \renewcommand{\theequation}{\arabic{section}.\arabic{equation}}%
  \renewcommand{\)}{\end{equation}}}

\newcommand{\aalpheqn}{\setcounter{saveeqn}{\value{equation}}%
  \stepcounter{saveeqn}\setcounter{equation}{0}%
  \renewcommand{\theequation}{\mbox{
        \Alph{subsection}.\arabic{saveeqn}\alph{equation}}}
   \renewcommand{\)}{\end{equation}}}
\newcommand{\areseteqn}{\setcounter{equation}{\value{saveeqn}}%
  \renewcommand{\theequation}{\Alph{subsection}.\arabic{equation}}%
  \renewcommand{\)}{\end{equation}}}

\renewcommand{\thefootnote}{\alph{footnote}}
\renewcommand{\(}{\begin{equation}}
\renewcommand{\)}{\end{equation}}
\newcommand{\ba}{\begin{eqnarray}}
\newcommand{\ea}{\end{eqnarray}}

\newcommand{\bp}{\mathop{\vtop{\ialign{##\crcr
   $\hfil\displaystyle{}\hfil$\crcr\noalign{\kern-13pt\nointerlineskip}
   \BIG{(}\hskip0pt\crcr\noalign{\kern3pt}}}}}
\newcommand{\cbp}{\mathop{\vtop{\ialign{##\crcr
   $\hfil\displaystyle{}\hfil$\crcr\noalign{\kern-13pt\nointerlineskip}
   \BIG{)}\hskip0pt\crcr\noalign{\kern3pt}}}}}
\newcommand{\pa}{\mathop{\vtop{\ialign{##\crcr

$\hfil\displaystyle{\oplus}\hfil$\crcr\noalign{\kern+1pt\nointerlineskip
}
   \hspace{.08in}$^{\alpha=0}$\hskip6pt\crcr\noalign{\kern3pt}}}}}

\newcommand{\Z}{\ensuremath{\mathbb Z}}

\newcommand{\beq}{\begin{equation}}
\newcommand{\eeq}{\end{equation}}

\numberwithin{equation}{section}

\catcode`\@=11
\def\vereq#1#2{\lower3pt\vbox{\baselineskip1.5pt \lineskip1.5pt
\ialign{$\m@th#1\hfill##\hfil$\crcr#2\crcr\sim\crcr}}}
\catcode`\@=12

\makeatletter
\newcommand\figcaption{\def\@captype{figure}\caption}
\newcommand\tabcaption{\def\@captype{table}\caption}
\makeatother
\renewcommand{\(}{\begin{equation}}
\renewcommand{\)}{\end{equation}}

\newcommand{\ZZ}{{\mathbb Z}}

\theoremstyle{plain}

\theoremstyle{definition}

\begin{document}

\begin{titlepage}
\begin{flushright}

hep-th/0501245
\end{flushright}

\vspace{2em}
\def\thefootnote{\fnsymbol{footnote}}

\begin{center}
{\Large\bf M-theory and characteristic classes}
\end{center}
\vspace{1em}

\begin{center}
Hisham Sati \footnote{E-mail: {\tt hsati@maths.adelaide.edu.au}\\
Research supported by the Australian Research Council. }
\end{center}

\begin{center}
\vspace{1em}
{\em {  Department of Physics\\
and \\
Department of Pure Mathematics\\
     University of Adelaide\\
       Adelaide, SA 5005,\\
       Australia}}\\
\end{center}

\vspace{0em}
\begin{abstract}
\noindent In this note we show that the Chern-Simons and the
one-loop terms in the M-theory action can be written in terms of new
characters involving the M-theory four-form and the string classes.
This sheds a new light on the topological structure behind M-theory
and suggests the construction of a theory of `higher'
characteristic classes.
\end{abstract}

\vfill

\end{titlepage}
\setcounter{footnote}{0}
\renewcommand{\thefootnote}{\arabic{footnote}}

\pagebreak
\renewcommand{\thepage}{\arabic{page}}

%%%%%%%%%%%%%%%%%%%%%%
\section{Introduction}
%%%%%%%%%%%%%%%%%%%%%

M-theory has emerged during the last decade as the theory that
unifies string theories and thus is the best candidate for a theory
of quantum gravity \cite{var} \cite{T} \cite{D}. This theory is
believed to be very rich both physically and mathematically.
However, such structures are only being discovered and a complete
description (let alone an understanding) of the theory remains a
grand challenge. The study of the partition function related to the
four-form has uncovered deep connections to K-theory \cite{DMW}, and
to a less-understood extent twisted K-theory \cite{relations} and
elliptic cohomology \cite{KS1}.

\vspace{3mm} In this note, we make further observations on the
topological part of the action, that we hope will provide an
approach that makes the topological structure more transparent. Our
arguements are rather intuitive but seem to point to some deep
mathematical theory. In particular, the structure of the action
motivated us to propose defining new ``higher'' characteristic
classes based on the Pontrjagin classes. We define the total string
class and the string character and we encode the action in terms of
a character built out of the four-form and the string classes.

\vspace{3mm} The M-theory action contains the usual
eleven-dimensional supergravity terms \cite{CJS}, namely the
Einstein-Hilbert, the $G_4$ kinetic terms, the Rarita-Schwinger
term, as well as the subtle topological terms, namely the
Chern-Simons term and the one-loop term \cite{Loop} given by (see
\cite{M} for a review)
\begin{eqnarray}
S_{11}&=&S_{CS} + S_{1-loop}
\nonumber\\
&=&\int_{Y^{11}} \frac{1}{6}C_3 \wedge G_4 \wedge G_4 -
\int_{Y^{11}} C_3 \wedge I_8(g)
\end{eqnarray}
where $[I_8(g)]=\frac{p_2 -(p_1/2)^2}{48}$, written in terms of the
Pontrjagin classes of the tangent bundle.

\vspace{3mm} Using cobordism, Witten has uncovered a structure
related to $E_8$ index theory by writing the above action on a
twelve-dimensional manifold $Z^{12}$ whose boundary is the M-theory
eleven-manifold $Y^{11}$. In twelve dimensions, the topological part
of the action is then
\begin{eqnarray}
S_{12}&=&\int_Z I_{12}
\nonumber\\
&=&\int_Z \frac{1}{6}G_4 \wedge G_4 \wedge G_4 -G_4 \wedge I_8
\end{eqnarray}
Of particular interest is the mod 6 congruence which was derived in
\cite{Al} using physical arguments related to Ho\v{r}ava-Witten
boundaries and in \cite{Wi1} using $E_8$ index theory.

%%%%%%%%%%%%%%%%%%%%%%
\section{The proposal}
%%%%%%%%%%%%%%%%%%%%%%
Here we propose writing the chern-Simons term in an exponentiated
form as \( \left[e^{G_4}\right]_{(12)} =\frac{1}{6}G_4 \wedge G_4
\wedge G_4, \label{expG}\) which looks like a character. Obviously, this 
very
simply encodes the correct normalization factor.

\vspace{3mm} Next we look at the one-loop term. Define the total
string class in terms of the individual string classes as \(
\lambda=\lambda_0 + \lambda_1 + \lambda_2 + \cdots \) where
$\lambda_0=1$, $\lambda_1=p_1/2$ is the usual string class, and we
define $\lambda_2$ to be $p_2/2$, as the second string class, and so
on. Define the exponentiated class (i.e. the ``total string
character''´´), $e^{\lambda}$, whose degree eight component
gives \( \frac{1}{2}\left( \lambda_1^2 -2 \lambda_2
\right). \label{8}\)
The one-loop gravitational polynomial $I_8$, rewritten in terms of the 
introduced string classes $\lambda_1$ and $\lambda_2$, is then exactly 
minus 24 times the eight-form (\ref{8}). The minus sign will be useful 
since $I_8$ shows up as in the action with the relative minus sign. 
The point out of this manipulation is that
this is analogous to writing the second Chern character in terms of
the Chern classes as \( ch_2=\frac{1}{2}(c_1^2 -2c_2). \)
%\vspace{3mm}
Going back to the total topological action, this can be written as \(
\left[e^{G_4}\left[1+\frac{1}{24}\left(e^{\lambda}-1\right)\right] 
\right]_{(12)}, \label{expression}\) the
degree twelve component. This apparently predicts a term \(
\frac{1}{2}G_4 \wedge G_4 \wedge \frac{\lambda_1}{24}, \) that can
be written in terms of $\sqrt{\widehat{A}}$ (without overall
factors) \footnote{In fact one can ask, in a different but related
context, whether one can replace the K-theoretic formula
$F(x)=ch(x)\sqrt{\widehat{A}(X)}$ for the RR fields \cite{MM}
\cite{MW} by something like $F(x)=-\frac{1}{24}ch(x)e^{\lambda}$ (note 
the
sign). They match for the lower two degrees of the gravitational
part (i.e. zero and 4), but not the degree 8. We find this
interesting but will not elaborate here on whether this is merely a
coincidence in low degrees or whether it leads to a correction of
the K-theoretic formula. The latter would be related to refinements
proposed in \cite{KS3}. We will explore this elsewhere.}, \( \left[
\sqrt{\widehat{A}}~ \right]_{(4)}=-\frac{1}{48}p_1 =-\frac{1}{24}\left[
e^{\lambda} \right]_{(4)}=\frac{1}{24}\left[
e^{-\lambda} \right]_{(4)}. \) We can find a rationale 
to exclude
such terms (as well as terms containing $\lambda_3$) on the basis of
parity. We want to retain the terms such that \(
e^{G_4}e^{\lambda}=-e^{-G_4}e^{\lambda} \) which kills the
terms with even number of $G_4$'s and keeps the ones with odd number
of $G_4$'s. So the total topological action is given by the
parity-odd part of the twelve-form component of the character.
\footnote{Note that Lagrangian of eleven-dimensional supergravity
has a symmetry given by reversing the sign of the 3-form potential
$C_3$, accompanied by a reversal in sign of odd number of space
coordinates \cite{DNP}. } Such parity conditions are in fact not so 
foreign since they show up in M-theory in the calculation of higher order 
corrections, including the one-loop term. However, mathematically, it does 
not seem that a priori we need to impose such conditions on 
(\ref{expG}), and later on (\ref{index}), in
order to talk about the cohomology theory below.

\vspace{3mm} The phase of the M-theory partition function \cite{DMW}
\cite{DFM} \cite{FM} would then be written as \( \Phi(C_3)=
(-1)^{\frac{1}{2}I_{R.S.}} \exp 2\pi i \left[\int_{Z^{12}}
e^{G_4}\left[1+\frac{1}{24}(e^{\lambda}-1)\right] \right], \) where it 
is understood that we
pick the degree twelve component of the integrand. Note the sign
ambiguity in the phase is canceled by the one coming from the phase
of the Pfaffian of the Rarita-Schwinger operator given in terms of
the Rarita-Schwinger action $I_{R.S.}$.

%%%%%%%%%%%%%%%%%%%%%%%%%
\section{A cohomology theory?}
%%%%%%%%%%%%%%%%%%%%%%%%%%
Here we draw an analogy to the construction of the Chern character
based on ordinary curvatures of connection one-forms. We will see
that this suggests that an analogous theory based on the Pontrjagin
classes (more precisely the string classes) instead of the Chern
classes seems to emerge.

\vspace{3mm} Recall the Chern character \( ch(F_2)=e^{F_2}=rk + F_2
+ \frac{1}{2}F_2 \wedge F_2 + \frac{1}{6} F_2 \wedge F_2 \wedge F_2
+ \cdots \) Now, in light of the above construction, we would like
to do the same for $G_4$ 
\footnote{Clearly, the Chern character
should be  written in terms of Lie-algebra valued curvature, whose
trace we suppress in an obvious way. The character corresponding to
$G_4$ does not seem to have such a valuedness since $G_4$ itself
does not correspond to structure group in a literal way. Of course,
it is true that if it is to be a (2-)gerbe then the simplest way is
to take it to be abelian and thus corresponding to a U(1) in an
appropriate sense. If one writes the C-field in terms of $E_8$ gauge
fields (e.g. take $C_3=CS_3(A)$ literally as in \cite{ES} or in a
more refined way as in \cite{DFM}) then $G_4$ might have some
nonabelian aspect to it. In any case, we leave such possibilities
open.}

\( e^{G_4}= c + G_4 + \frac{1}{2}G_4 \wedge G_4 + \frac{1}{6} G_4
\wedge G_4 \wedge G_4 + \cdots \) which we would like to think of as
a sort of M-theoretic character ${\mathbb M}$, analogous to the
Chern character. The term "c" should refer to the appropriate
concept in this case that replaces the rank of the bundle for the
case of K-theory. \footnote{We do not see any reason for the
constant term to be different from one, otherwise the normalizations are
altered. However, the formalism can accomodate other possibilities.}

\vspace{3mm} The $E_8$ and Rarita-Schwinger indices are then encoded
in \footnote{In the definition of ${\mathbb M}$ we have to also
insert the gravitational correction term containing $e^{\lambda}$ to be 
able to get the total action and thus for the statement to be correct. We
however look only at $G_4$ at this stage and adding the
gravitational term is straightforward. We simply replace ${\mathbb
M}$ by ${\mathbb M}'={\mathbb 
M}\left[1+\frac{1}{24}\left(e^{\lambda}-1\right)\right]$. Again, exclusion 
of some terms can be based on parity: odd (even) parity for odd-
(even-)rank character.} \( {\rm Index} (\rm M~object)=\int {\mathbb M} =
\int e^{G_4}. \label{index} \) The individual components are
\begin{eqnarray}
{\mathbb M}_0 &=& 1
\nonumber\\
{\mathbb M}_1 &=& G_4
\nonumber\\
{\mathbb M}_2 &=& \frac{1}{2} G_4 \wedge G_4
\nonumber\\
{\mathbb M}_3 &=& \frac{1}{6} G_4 \wedge G_4 \wedge G_4,
\end{eqnarray}
the latter three corresponding to the M2-brane, the M5-brane/little
M-theory, and M-theory respectively (with the correct
normalization!). We see that as a byproduct the M-branes and
``M-theories'' receive a somewhat unified treatment.

\vspace{3mm} Let us explain this. The Chern-Simons construction for
type IIA string theory and for the M-fivebrane go in a very similar
way. In both we have a manifold of dimension $4k+2$ with $k=1$ for
the fivebrane and $k=2$ for type IIA string theory. The Chern-Simons
construction requires extending the $(4k+2)$-dimensional manifold
$X$ to a $(4k+3)$-dimensional manifold $X\times S^1$ by an extra
$S^1$, which, in the case of type IIA is just the extension to
M-theory. Then the construction requires extending the resulting
manifold to a $(4k+4)$-dimensional ``coboundary'', i.e. an $N$ such
that $\partial N=X \times S^1$. In the case of IIA/M-theory, this is
just the manifold $Z$ we encountered in the introduction. In both
cases one is extending the manifold together with a four-class, so
that this requires the vanishing of the spin cobordism cohomology
group $MSpin_{4k+3}(K(\ZZ,4))$. This was shown to be the case by
Stong for $k=2$ \cite{St} and by Hopkins-Singer for $k=1$
\cite{Hop}. Now, similarly to the case of the Chern-Simons term in
M-theory, we propose writing the corresponding quadratic term
involving only $G_4$ on $N_8$ in a similar exponential way, i.e

\(
\left[ e^{G_4}\right]_{(8)}=\frac{1}{2} G_4 \wedge G_4
\)

which is the formula derived by Witten \cite{Five} and also in a
more general situation by Hopkins-Singer \cite{Hop}. \footnote{The
two references differ by an overall minus sign, as pointed out in
\cite{Hop}.} \footnote{There is another term in the fivebrane action
extended to 8-dimensions. This is $G_4 \wedge \lambda_1$. To include
this term it seem that one has to look at the combination
$e^{G_4}e^{\lambda}$ in this case.} Another way of looking at the 
fivebrane in eleven dimensions is using tubular neighborhoods and disk 
bundles as in \cite{Harvey}.

\vspace{3mm}
The above suggests that there is a generalized
cohomology theory in which $G_4$ lives and the character is a
multiplicative map from the theory of M-objects to $4k$-th
cohomology. \( {\mathbb M}=e^{G_4}: {\mathcal M} \to H^{4k} \) where
${\mathcal M}$ is the (generalized cohomology?) theory that
describes M-theory.

\vspace{3mm}
Recall that the Chern character is a map from K-theory
to even cohomology \( ch: K \to H^{ev} \) and satisfies (for two
vector bundles $E$ and $F$)
\begin{eqnarray}
ch(E\oplus F)=ch(E) \oplus ch(F)
\\
ch(E \otimes F)=ch(E) \wedge ch(F).
\end{eqnarray}

For ${\mathcal E}$ and ${\mathcal F}$ ``M-objects'', we want
${\mathbb M}$ to have properties analogous to those of the Chern
character, i.e.
\begin{eqnarray}
{\mathbb M}({\mathcal E}\oplus {\mathcal F})= {\mathbb M}({\mathcal
E}) \oplus {\mathbb M}({\mathcal F})
\nonumber\\
{\mathbb M}({\mathcal E}\otimes {\mathcal F})= {\mathbb M}({\mathcal
E}) \wedge {\mathbb M}({\mathcal F}).
\end{eqnarray}
Note further that $d{\mathbb M}=0$ since $G_4$ itself is closed by
virtue of the Bianchi identity.

%%%%%%%%%%%%%%%%%%%%%
\vspace{3mm} We have proposed a theory of multiplicative classes based on 
the Pontrjagin classes (more precisely, the string classes, the higher 
ranks of which we defined above) instead of the usual Chern characters 
which are built out of the Chern classes.  
For K-theory, the objects corresponding to the two-form
curvature $F_2$ are vector bundles. What are the corresponding
M-objects, i.e. the ones related to $G_4$? We do not have a precise 
answer but we propose that they would have to do with `2-objects', e.g. 
2-gerbes or more precise refinements as in \cite{DFM}. This will be 
discussed  seperately.

\vspace{3mm} What is the theory we are looking for? From the general
structure of the character, from the mod 24 congruence of the string
character, and from the previous work \cite{KS1} \cite{KS2} \cite{KS3},
we expect such a theory to be some form/refinement of elliptic
cohomology (e.g. related to the theory of topological modular
forms).  From the mathematical point of view, in order to have a
theory of {\it characteristic} classes, we need to have a classifying 
space from which we pullback structures to our spaces. This, and 
the relation to the $E_8$ and Rarita-Schwinger
bundles, will also be explored seperately.

%\vspace{3mm}
%We hope to make our intuitive observations and arguements
%in this note
%more mathematically rigorous in the future.

\vspace{3mm} {\bf Acknowledgements}

\vspace{2mm} The author would like to thank Igor Kriz for comments
on the manuscript and the Michigan Center for Theoretical Physics
for hospitality. The author is also indebted to Greg Moore for
interesting questions, discussions and remarks, and in particular for 
pointing out an anomalous factor in the earlier version, and for proposing 
its remedy.

%\newpage

%%%%%%%%%%%%%%%%%%%%%%%%%%%%%

%


\begin{thebibliography}{99}
%%%%%%%%%%%%%%%%%%%%%%%%%%%
\bibitem{var}
E.~Witten, {\it String theory dynamics in various dimensions},
Nucl.~Phys. {\bf B443} (1995) 85, [{\tt arXiv:hep-th/9503124}].

\bibitem{T}
P.~K.~Townsend, {\it Four lectures on M-theory}, [{\tt
arXiv:hep-th/9612121}].

\bibitem{D}
M.~J.~ Duff (ed.), {\it The World in eleven dimensions :
Supergravity, supermembranes and M-theory}, Institute of Physics
Publishing, Bristol, (1999).

\bibitem{DMW}
E. Diaconescu, G. Moore and E. Witten, {\it $E_8$ gauge theory, and
a derivation of K-theory from M-theory} Adv. Theor. Math. Phys. {\bf
6} (2003) 1031, [{\tt arXiv:hep-th/0005090}].

\bibitem{relations}
V. Mathai and H. Sati, {\it Some relations between twisted K-theory
and $E_8$ gauge theory}, J. High Energy Phys. {\bf 0403} (2004) 016,
[{\tt arXiv:hep-th/0312033}].

\bibitem{KS1}
I.~Kriz and H.~Sati, {\it M Theory, type IIA superstrings, and
elliptic cohomology}, Adv. Theor. Math. Phys. {\bf 8} (2004) 345,
[{\tt arXiv:hep-th/0404013}].

\bibitem{CJS}
E.~Cremmer, B.~Julia and J.~Scherk, {\it Supergravity theory in
eleven dimensions}, Phys. Lett. {\bf B76} (1978) 409.

\bibitem{Loop}
M. J. Duff, James T. Liu and R. Minasian, {\it Eleven dimensional
origin of string/string duality: A one loop test}, Nucl. Phys. {\bf
B452} (1995) 261, [{\tt arXiv:hep-th/9506126}].

\bibitem{M}
G. Moore, {\it Anomalies, Gauss laws, and Page charges in M-theory},
[{\tt arXiv:hep-th/0409158}].

\bibitem{Al}
S. P. de Alwis, {\it Anomaly cancellation in M-theory}, Phys. Lett.
{\bf B392} (1997) 332, [{\tt arXiv:hep-th/9609211}].


\bibitem{Wi1}
E. Witten, {\it On flux quantization in M-theory and the effective
action}, J. Geom. Phys. {\bf 22} (1997) 1, [{\tt
arXiv:hep-th/9609122}].

\bibitem{MM}
R. Minasian and G. Moore, {\it K-theory and Ramond-Ramond charge},
J. High Energy Phys. {\bf 11} (1997) 002, [{\tt
arXiv:hep-th/9710230}].

\bibitem{MW}
G.~Moore and E.~Witten, {\it Self duality, Ramond-Ramond fields, and
K-theory}, J. High Energy Phys. {\bf 05} (2000) 032, [{\tt
arXiv:hep-th/9912279}].

\bibitem{KS3}
I.~Kriz and H.~Sati, {\it Type II string theory and modularity},
[{\tt arXiv:hep-th/0501060}].

\bibitem{DNP}
M. Duff, B. Nilsson and C. Pope, {\it Kaluza-Klein supergravity},
Phys. Rept. {\bf 130} (1986) 1.

\bibitem{DFM}
E. Diaconescu, D. S. Freed and G. Moore, {\it The M-theory 3-form
and $E_8$ gauge theory}, [{\tt arXiv:hep-th/0312069}].

\bibitem{FM}
D. S. Freed and G. Moore, {\it Setting the quantum integrand of
M-theory}, [{\tt arXiv:hep-th/0409135}].

\bibitem{ES}
J. Evslin and H. Sati, {\it SUSY vs $E_8$ gauge theory in 11
dimensions}, J. High Energy Phys. {\bf 0305} (2003) 048, [{\tt
arXiv:hep-th/0210090}].

\bibitem{St}
R. E. Stong,
 {\it Calculation of $\Omega^{{\rm Spin}}_{11}(K(\Z,4))$}, in
 Workshop on Unified String Theories, World
Sci. Publishing, Singapore (1986) 430.

\bibitem{Hop}
M. J. Hopkins and I. M. Singer, {\it Quadratic functions in
geometry, topology,and M-theory}, [{\tt arXiv:math.AT/0211216}].

\bibitem{Five}
E. Witten, {\it Five-brane effective action In M-theory}, J. Geom.
Phys. {\bf 22} (1997) 103, [{\tt arXiv:hep-th/9610234}].

\bibitem{Harvey}
D. Freed, J. A. Harvey, R. Minasian and G. Moore,{\it
Gravitational anomaly cancellation for M-theory fivebranes},
Adv. Theor. Math. Phys. {\bf 2} (1998) 601,
[{\tt arXiv:hep-th/9803205}].

\bibitem{KS2}
I.~Kriz and H.~Sati, {\it Type IIB string theory, S-duality and
generalized cohomology}, [{\tt arXiv:hep-th/0410293}].

\end{thebibliography}
\end{document}